\documentclass[twocolumn,showpacs, preprintnumbers, amsmath, amssymb, superscriptaddress,pra]{revtex4}
\usepackage{graphicx}
\usepackage{dcolumn}
\usepackage{bm}
\usepackage{epsf}
\usepackage{amssymb}
\DeclareGraphicsExtensions{.eps}
\usepackage{epstopdf}
\usepackage{natbib}
\usepackage{subfigure}
\usepackage{graphicx}
\begin{document}

\author{David S. Simon}
\affiliation{Dept. of Electrical and Computer Engineering, Boston
University, 8 Saint Mary's St., Boston, MA 02215}
\affiliation{Dept. of Physics and Astronomy, Stonehill College, 320 Washington Street, Easton, MA 02357}
\author{Casey Fitzpatrick}
\affiliation{Dept. of Electrical and Computer Engineering, Boston
University, 8 Saint Mary's St., Boston, MA 02215}
\author{Alexander V. Sergienko}
\affiliation{Dept. of Electrical and Computer Engineering, Boston
University, 8 Saint Mary's St., Boston, MA 02215}
\affiliation{Photonics Center, Boston
University, 8 Saint Mary's St., Boston, MA 02215}
\affiliation{Dept. of Physics, Boston University, 590 Commonwealth
Ave., Boston, MA 02215}

\begin{abstract}
We propose an interferometric method for statistically discriminating between nonorthogonal states in high dimensional Hilbert spaces for use in
quantum information processing. The method is illustrated for the case of photon orbital angular momentum (OAM) states.  These states belong to pairs of bases that are mutually unbiased on a sequence of two-dimensional subspaces of the
full Hilbert space, but the vectors within the same basis are not necessarily orthogonal to each other. Over multiple trials, this method allows distinguishing
OAM eigenstates from superpositions of multiple such eigenstates. Variations of the same method are then shown to be capable of preparing and detecting arbitrary
linear combinations of states in Hilbert space. One further variation allows the construction of chains of states obeying recurrence relations on the Hilbert
space itself, opening a new range of possibilities for more abstract information-coding algorithms to be carried out experimentally in an simple manner. Among other applications, we show that this approach provides a simplified means of switching between pairs of high-dimensional mutually unbiased OAM bases.
\end{abstract}

\title{Discrimination and Synthesis of Recursive Quantum States in High-Dimensional Hilbert Spaces}


\maketitle


\section{Introduction}\label{introsection}

Quantum key distribution (QKD) aims to enable two agents, Alice and Bob, to generate a shared cryptographic key while preventing an unauthorized eavesdropper,
Eve, from gaining significant information about the key without being revealed. In the context of optics, the most common variable used to encode the key is the
photon's polarization. Polarization spans a two-dimensional space, so that it normally can encode only one bit of key per photon. In order to generate multiple
bits per photon, other variables may be employed which span higher dimensional Hilbert spaces \cite{englund,chen}. The most promising of these is the photon's
orbital angular momentum (OAM). The OAM along the propagation axis is $L_z=l\hbar$, where the topological charge $l$ is restricted to integer values, and in
principle can be arbitrarily large. If a stream of photons is produced whose $l$ values are allowed to range between $-L$ and $+L$, then each photon can encode
up to $\log_2(2L+1)$ bits of information in its OAM.

In \cite{simon1}, an approach was introduced for using OAM to generate a secret encryption key based on a novel entangled light source
\cite{liew,trevino,trevino2,dalnegro,lawrence} that produces output with an OAM spectrum whose absolute values are restricted to the Fibonacci sequence. Recall that
the Fibonacci sequence starts with initial values $F_1=1$ and $F_2=2$, generating the rest of the sequence via the recurrence relation $F_{n+2}=F_{n+1}+F_n$.
These states are used to pump a nonlinear crystal, leading to spontaneous parametric down conversion (SPDC), in which a small proportion of the input
photons are split into two entangled output photons, called the signal and idler. After the crystal, any photons with non-Fibonacci values of OAM are filtered
out. The result is that angular momentum conservation and the Fibonacci recurrence relation force the two output photons to have OAM values that are adjacent
Fibonacci numbers (for example $F_m$ and $F_{m+1}$). These photons can then be used to generate a secret key; see \cite{simon1,simon2} for more details on the procedure.

The Fibonacci key distribution protocol requires the ability to distinguish between single OAM eigenstates and superpositions of pairs of eigenstates. These two
types of states belong to two mutually unbiased bases on subspaces of the full Hilbert space. The state discrimination cannot be done unambiguously on a single
trial, just as it can not be determined in a single trial whether a photon's polarization is vertical or diagonal; however, over many trials a statistical
picture of the outcomes can be built up in such a way that eavesdropping alters the outcome probability distributions in a detectable manner. In this manner, an
eavesdropper can be revealed over multiple trials even though it may not be possible to identify errors on any individual trial. The details of how the
eavesdropper revelation works may be found in \cite{simon2}.

The Fibonacci protocol is an example of a more abstract approach to QKD in which, rather than simply modulating between two spatial measurement bases,
more abstract modulations are carried out, switching between more general sets of states in Hilbert space. Unlike polarization-based protocols, the relevant
unbiased bases in Hilbert space do not correspond directly to projections onto bases in physical space, allowing more freedom in the choice of bases used. By
opening up a broader range of states to manipulate, this approach has promise to increase the key capacity per photon, to allow new methods for safeguarding the
security of the key, and in some cases to simplify experimental implementations. In order to carry out this program, it is necessary to have a simple means of
producing and detecting general (possibly nonorthogonal) linear combinations of basis states.

In this paper, we examine the issue of discriminating between non-orthogonal states formed from superpositions of OAM eigenstates. We propose an interferometric
method which allows the user to distinguish between two different superpositions or to distinguish superpositions from individual eigenstates. There is some
probability of error in the state identification, due to the nonorthogonal nature of the states involved, but for the purposes of QKD this can be dealt with in
the reconciliation stage, as detailed in \cite{simon2}. More general discussions of the problem of non-orthogonal state discrimination with minimal error may be
found in \cite{helstrom,bergou,cheffles}. Another approach to the study of superposition states in OAM space via optical filtering with spatial light modulators
can be found in \cite{jack2,radwell}. In \cite{bussieres}, an approach was taken to high-dimensional time-bin-encoded states that is similar in spirit to what is
done here, allowing measurements to be done in arbitrary bases.

We first display a solution to the problem of sorting pairwise superpositions of states. We then generalize the apparatus to allow the detection of arbitrary
linear superpositions of OAM states.  It is further shown how slight changes could turn the same structure into a means of \emph{synthesizing} arbitrary
superposition states. This is similar to the case of conventional phase and polarization coding in telecom systems, where Mach-Zehnder interferometers are used
for both coding and decoding of the signal. In the current case, the type of tree-like structure used to generate and detect these superposition states will be
referred to as a \emph{superposition generation and detection tree} (SGDT). The use of different SGDT's allows the generalization of the Fibonacci protocol to similar
protocols based on other linear recurrence relations. Finally, we move from recurrence relations on the real numbers to recurrence relations on Hilbert space,
showing how to construct chains of \emph{states} (as opposed to eigenvalues) that obey arbitrary linear recurrence relations; this is done via a set of nested
structures we will refer to as a recursive state generator (RSG).

This paper is organized as follows. In section \ref{setupsection} we review the Fibonacci protocol as a motivating example for what follows. In section
\ref{supersection}, we then discuss the problem of nonorthogonal state discrimination and introduce SGDT's for the states relevant to the Fibonacci protocol,
with generalizations that cover other types of superposition states in section \ref{generalsection}. We then introduce RSG's in section
\ref{stategensection} for the preparation of superposition states, and discuss the idea of recurrent sets of states in Hilbert space. Finally, we briefly discuss
conclusions in section \ref{conclusionsection}.

\begin{figure}
\begin{center}
\includegraphics[scale=.28]{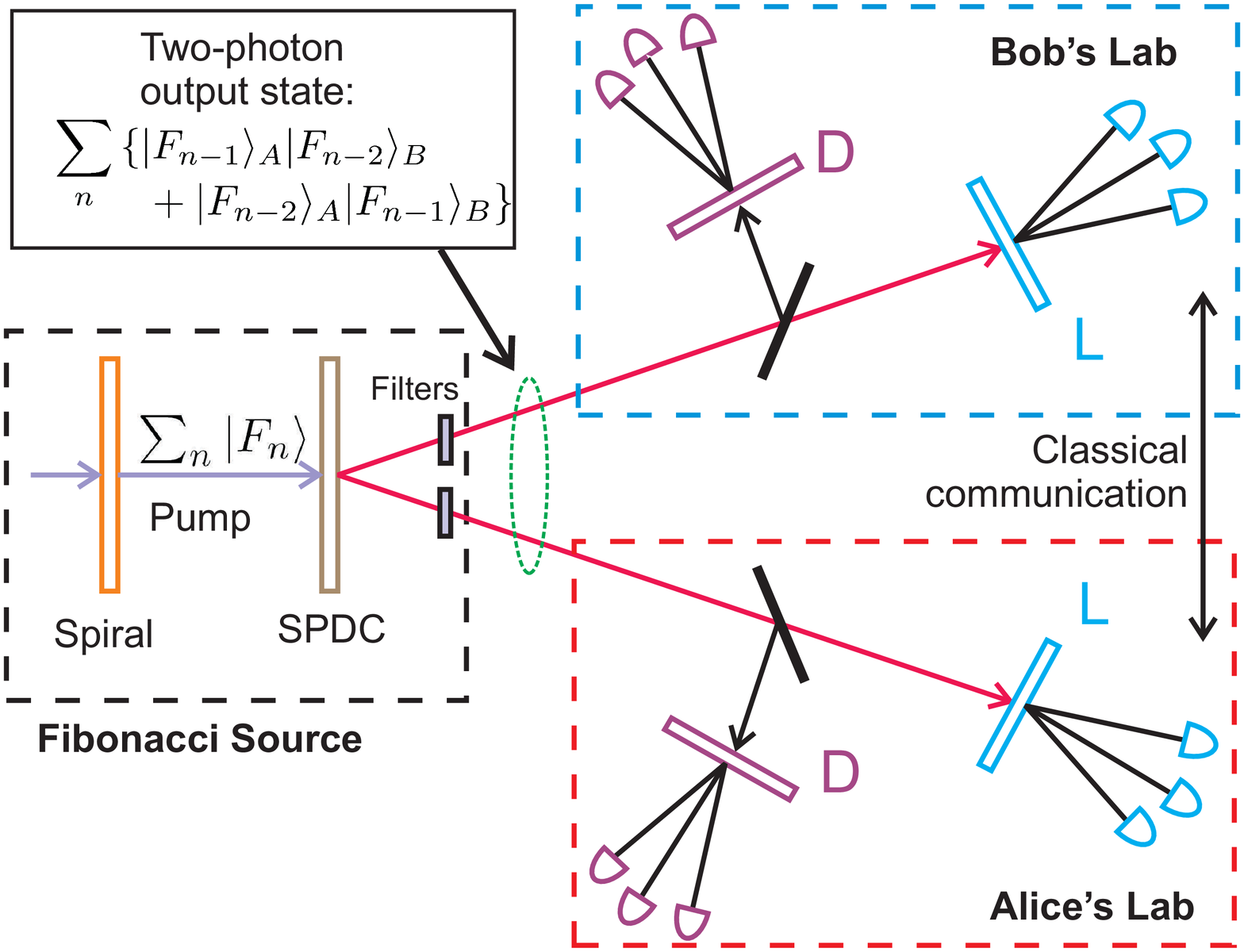}
\caption{(Color online) Schematic setup for generating cryptographic key with Fibonacci-valued OAM states. The $\mathbb{L}$ and
$\mathbb{D}$ detection stages respectively test for different OAM eigenstates or different superposition states. Interference between
scatterings in aperiodic Vogel spiral lead to an OAM spectrum in the pump that is restricted to the Fibonacci sequence. The filters after
the crystal remove any non-Fibonacci values that appear in the down conversion process. See \cite{simon1}
for more detailed description of the apparatus.}
\label{setupfig}
\end{center}
\end{figure}

\section{Fibonacci Key Distribution}\label{setupsection}

To provide a motivating example of the need for distinguishing between nonorthogonal superposition states within unbiased bases, we review the essential ideas of
the Fibonacci protocol \cite{simon1}, which allows multiple key bits to be generated per photon. In this protocol, distinguishing such states is necessary both
to determine the key and in order to provide security against eavesdropping. Security of the protocol stems from the random switching between measurements in two
bases (the $\mathbb{D}$ and $\mathbb{L}$ bases introduced below) that are mutually unbiased on each of a chain of two dimensional subspaces; but in one of these
bases the two basis vectors are mutually nonorthogonal.

The basic setup of \cite{simon1} is shown schematically in Fig. \ref{setupfig}. The source on the left consists of an aperiodic nano-array of scatterers in the
form of a Vogel spiral \cite{liew,trevino,trevino2,dalnegro,lawrence}, followed by a nonlinear down conversion crystal. The Vogel spiral causes the scattered
waves from the different scattering centers to interfere in such a way that only components with OAM belonging to the Fibonacci sequence survive.  As a result
the pump beam entering the crystal is a superposition of Fibonacci-valued OAM states, $\sum_n|F_n\rangle$. After the crystal, the signal and idler states have
OAM values that must sum to a Fibonacci number, but are not necessarily Fibonacci numbers themselves. Filters in the signal and idler paths then remove the
non-Fibonacci values. The result is that at the output of the source there is an entangled two-photon state with Fibonacci-valued OAM in the two output
directions:
\begin{equation}\psi =\sum_n \big( |F_{n-1}\rangle_A  |F_{n-2}\rangle_B + |F_{n-2}\rangle_A |F_{n-1}\rangle_B\big) ,\label{initialstate}\end{equation} where the
index $n$ runs over the indices of the allowed Fibonacci numbers in the pump beam: $|\Psi\rangle_{pump} =\sum_n |F_n\rangle.$  Alice and Bob each receive half of
the entangled pair. Each of them uses a  50/50 nonpolarizing beam splitter to randomly direct the photon either to one of two types of detection stages, referred
to as $\mathbb{L}$ and $\mathbb{D}$ stages. $\mathbb{L}$-type detection consists of an OAM sorter \cite{leach,berkhout,lavery} followed by a set of single-photon
detectors.  OAM sorters send different $l$ values into different outgoing directions, so that they register in different detectors, allowing the value of $l$ to
be determined. In contrast,  $\mathbb{D}$-type detection is used to distinguish between different superposition states $|S_n\rangle $ of the form
\begin{eqnarray}|S_n\rangle ={1\over \sqrt{2}}\big( |F_{n-1}\rangle+|F_{n+1}\rangle \big) .\end{eqnarray} The detection of
such superposition states can be accomplished in several ways \cite{miyamoto,jack,kawase}. One complication with the $\mathbb{D}$-type detection is that adjacent
superposition states, such as ${1\over \sqrt{2}}\left( |F_{n-1}\rangle+|F_{n+1}\rangle \right) $ and ${1\over \sqrt{2}}\left( |F_{n+1}\rangle+|F_{n+3}\rangle
\right) $ are not orthogonal, meaning that the two states cannot be unambiguously distinguished from each other. It is also necessary to keep the possible key
values uniformly distributed, so that Eve can't obtain any advantage from knowledge of the nonuniform distribution. These complications add some complexity to
the classical exchange (see \cite{simon2} for details) and alter the corresponding detection probabilities; however this extra complexity is well compensated by
the increased key-generating capacity. Moreover, the degree of complication does not grow with the size of the alphabet used, so that the benefits outweigh the
complications by a larger amount as the range of $l$ values increases.

Note that the action of the spiral does not lead to loss of any photons or energy. The energy is simply being redirected from non-Fibonacci to Fibonacci modes
via constructive and destructive interference, causing no reduction of photon efficiency in the apparatus. The signal and idler filters after the crystal, on the
other hand, do lower the efficiency through photon loss. The percent of photons retained is simply given by the fraction of the values in the chosen operating
range that fall on the Fibonacci sequence; for example if the eight Fibonacci values between $2$ and $55$ are used, this about $15\%$. Since no losses occur in
the spiral, this is not enough to lower the event rate below reasonable levels. As in all entanglement-based QKD protocols, the principal constraints on key
generation rate come simply from the low efficiency of the down conversion process itself and from losses in propagation between the source and receivers.

Assume that the pump spectrum is broad enough to be approximately flat over a
sufficient span to produce signal and idler OAM values of uniform probability over the range $F_{m_0}$ to $F_{m_0+N-1}$, for some $m_0$. The outcomes for
$\mathbb{L}$-type detection (OAM eigenstates) that will be used for key generation by Alice and Bob are then simply $|F_{m_0}\rangle ,|F_{m_0+1}\rangle, \dots
, |F_{m_0+N-1}\rangle$.
The outcomes for $\mathbb{D}$-type detection (two-fold OAM superposition states) used by Alice and Bob for key generation run from \begin{equation}
|S_{m_0}\rangle = {1\over \sqrt{2}}\left\{ |F_{m_0-1}\rangle +|F_{m_0+1} \rangle\right\} \end{equation} to
 \begin{equation} |S_{m_0+N-1}\rangle = {1\over \sqrt{2}}\left\{ |F_{m_0+N-2}\rangle +|F_{m_0+N} \rangle\right\}  .\end{equation}

\begin{figure}
\begin{center}
\includegraphics[scale=.28]{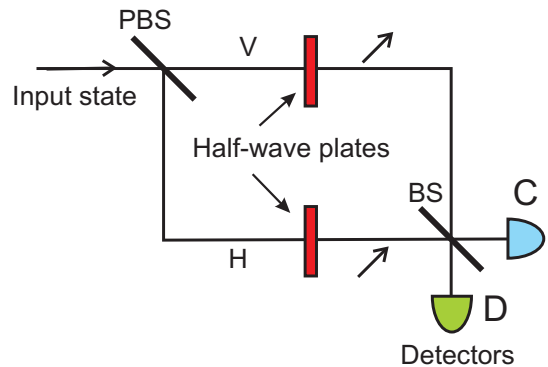}
\caption{(Color online) Apparatus for statistically distinguishing sets of vertically or horizontally polarized photons from diagonally-polarized photons.
The half-wave plates rotate vertical and horizontal polarization states to a diagonal state, in order to restore indistinguishability between paths.
If the photons are vertically or horizontally polarized, the two detectors $C$ and $D$ will register the same number of events. For
diagonally polarized states at $+45^\circ$ from the horizontal,
destructive interference will prevent detector $D$ from registering any events, with constructive interference occurring at detector $C$.
For inputs states polarized along the other diagonal ($-45^\circ$) the roles of the detectors are reversed: constructive interference occurs at $D$ and
destructive interference
prevents $C$ from firing. }
\label{polarfig}
\end{center}
\end{figure}

\section{Discriminating superposition states}\label{supersection}

The sorting of the OAM eigenstates $|F_n\rangle$ in the $\mathbb{L}$ basis is straightforward, but the sorting of the superposition states $|S_n\rangle$ in the
$\mathbb{D}$ basis is more problematic: because these states are not orthogonal with the neighboring states two units above and below them ($\langle S_{n\pm
2}|S_n\rangle ={1\over 2}$) they can not be distinguished unambiguously from each other. The complications this brings are the price of increasing the size of the coding space.

\begin{figure}
\begin{center}
\includegraphics[scale=.28]{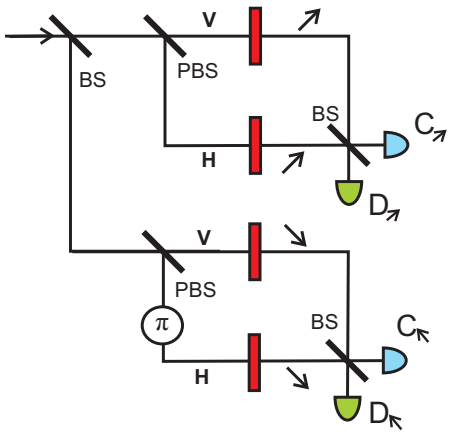}
\caption{(Color online) Two copies of the previous interferometer (Fig. \ref{polarfig}) can be combined, with the half-wave plates in the upper and lower copies taking the polarization vectors to opposite diagonals. If either of the states $|V\rangle $ or $|H\rangle$ is input, all four detectors have equal probability of firing. But if one of the two
superposition states $|\nearrow\rangle ={1\over
\sqrt{2}}\left( |\uparrow\rangle +|\rightarrow\rangle\right)$ or $|\nwarrow\rangle ={1\over
\sqrt{2}}\left( |\uparrow\rangle -|\rightarrow\rangle\right)$ is input, then only the $C$ detector in one interferometer and the $D$ detector in the
other can fire. (The $\pi$ phase shift in one branch of
the bottom interferometer is not necessary, but is inserted to keep the labeling of the $C$ and $D$ detectors consistent in the upper
and lower interferometers. )}
\label{polar2fig}
\end{center}
\end{figure}

To deal with this, we work by analogy to superposition states of polarization. Suppose Alice is sending a stream of photons to Bob, and that each photon is
either polarized in the vertical/horizontal basis or in the diagonal basis. Imagine that Bob sends each photon he receives through a Mach-Zehnder interferometer,
as shown in Fig. \ref{polarfig}. In this interferometer, a polarizing beam splitter (PBS) transmits vertically polarized photons and reflects horizontally
polarized photons. After separating the two polarization components, the each component is rotated by a half-wave plate so that the two polarization vectors
point along the same diagonal. It is then impossible to determine which path was taken to reach the second (nonpolarizing) beam splitter (BS). The amplitudes in
the two arms of the interferometer are then recombined and sent to the two detectors labeled $C$ and $D$. If the incoming photon was either vertically or
horizontally polarized, then the two detectors are equally likely to fire. On the other hand, if the initial photon was diagonally polarized, $|\nearrow\rangle
={1\over \sqrt{2}}\left( |\uparrow\rangle +|\rightarrow\rangle\right)$, then at the final beam splitter there will be interference between the two amplitudes. In
particular, there will be constructive interference at detector $C$ and destructive interference at detector $D$.

We may expand the apparatus of Fig. \ref{polarfig} to make its appearance more symmetrical and to make the analogy to the OAM case clearer below. By using a beam
splitter to couple the interferometer of Fig. \ref{polarfig} to a similar one in which the half-wave plates rotate polarizations to the opposite diagonal (Fig.
\ref{polar2fig}), we now have a setup in which vertically or horizontally polarized photons are equally likely to trigger any of the four detectors, but the
diagonally polarized superposition state $|\nearrow\rangle ={1\over \sqrt{2}}\left( |\uparrow\rangle +|\rightarrow\rangle\right)$ can only trigger detectors
$C_\nearrow$ and $D_\nwarrow$. Similarly, superposition state $|\nwarrow\rangle ={1\over \sqrt{2}}\left( |\uparrow\rangle -|\rightarrow\rangle\right)$ can only
trigger detectors $C_\nwarrow$ and $D_\nearrow$. On a single trial, it is impossible to determine the incoming polarization state of an individual photon, but
over many trials the distribution of the states received by Bob can be built up and compared to the distribution sent by Alice. This setup can therefore be used
to statistically detect tampering with the states en route.

\begin{figure}
\begin{center}
\includegraphics[scale=.3]{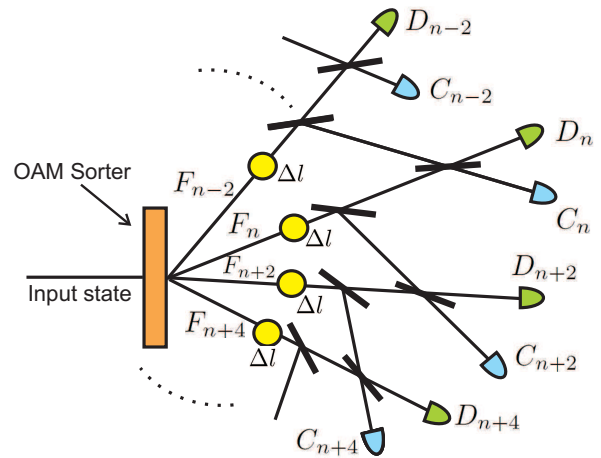}
\caption{(Color online) A portion of a detection unit for statistically detecting superpositions of OAM states, analogous to the polarization version of
the previous figure. The sorter separates different OAM values,
which are then shifted to zero OAM by spiral wave plates or holograms (the yellow circles).  From the top downward, the OAM shifters shown
change the incoming OAM value by $\Delta l=-F_{n-2}, -F_{n}, -F_{n+2},  -F_{n+4}$. Superpositions of the form $|F_n\rangle +|F_{n+2}\rangle$
cause constructive interference at the $C_n$
detectors and destructive interference at the $D_n$ detectors, while OAM eigenstates lead to equal detection rates at both types.
Coming out of the same sorter is a similar arrangement (not shown) for the states $\dots |F_{n-1}\rangle , \; |F_{n+1}\rangle , \; |F_{n+3}\rangle ,
\dots $} \label{CDfig}
\end{center}
\end{figure}

A similar idea can be used for OAM (Fig. \ref{CDfig}). An OAM sorter replaces the polarizing beam splitter, and a \emph{pair} of detectors $C_n$ and $D_n$ is
used at the output ports of the final nonpolarizing beam splitters. If $C_n$ fires during the key-generating trials, we count that as an $|S_{n-1}\rangle $
detection. Due to destructive interference, $D_n$ should not fire for superposition state input of the considered form, so its firing will count as an $|F_n\rangle $ detection. Then the scheme of ref. \cite{simon2} is used to reconcile Alice's and Bob's trials by classical information exchange in order to arrive at
an unambiguously agreed-upon key. During the security checks, the distribution of counts in $C_n$ and $D_n$ separately are examined in order to detect
eavesdropper-induced deviations from the expected probability distributions. In order to achieve the indistinguishability required for interference, the OAM of
each photon is shifted to zero after the sorting (by means of a spiral phase plate, for example). This is analogous to the use of a diagonal polarizer to restore indistinguishability in the polarization case. Note that measurements with detectors $C_n$ and $D_n$,
respectively, are equivalent to looking for nonzero projections onto the states
\begin{eqnarray}|C_n\rangle &=& {i\over \sqrt{2}} \left( |F_n\rangle +|F_{n-2}\rangle \right)  \; =\; i|S_{n-1} \rangle \\
|D_n\rangle &=& {1\over \sqrt{2}} \left( |F_n\rangle -|F_{n-2}\rangle \right) .
\end{eqnarray} These two sets of states are also mutually nonorthogonal; for equal $n$, we find $\langle C_n|D_n\rangle =0$, but more
generally $\langle C_n|D_m\rangle ={1\over 2}\left( \delta_{m,n-2}+\delta_{m,n+2}\right) .$ Note that $|C_n\rangle $ and $|D_n\rangle $
are the analogs of diagonal polarization states in the two-dimensional subspace spanned by $|F_n\rangle $ and $|F_{n-2}\rangle $.

The configuration of Fig. \ref{CDfig} is analogous to multiple copies of the interferometer of Fig. \ref{polarfig}, all being fed by a single OAM sorter. This generalizes Fig. \ref{polar2fig} from two interferometers to many. The
sorter plays the role of the polarizing beam splitter in Fig. \ref{polarfig}, directing different input states into different paths through the system.
Similarly, the diagonal polarizers and the OAM shifters play the same role in the two devices, each being used to restore indistinguishability to the amplitudes
that followed different paths, allowing these amplitudes to interfere when recombined at the detectors.  Eigenstates are equally likely to trigger the $C$ and
$D$ detectors, while superposition states of the form of $|S_n\rangle $ should never trigger the $D$ detectors.  As in the polarization case, the state of an
individual trial can't be determined, but the statistical distributions of detections are changed for different input states; the sole exception to this is that
if a $D$ detector fires we know that the state must be an eigenstate.


\section{Generalizations}\label{generalsection}

\begin{figure}
\begin{center}
\includegraphics[scale=.3]{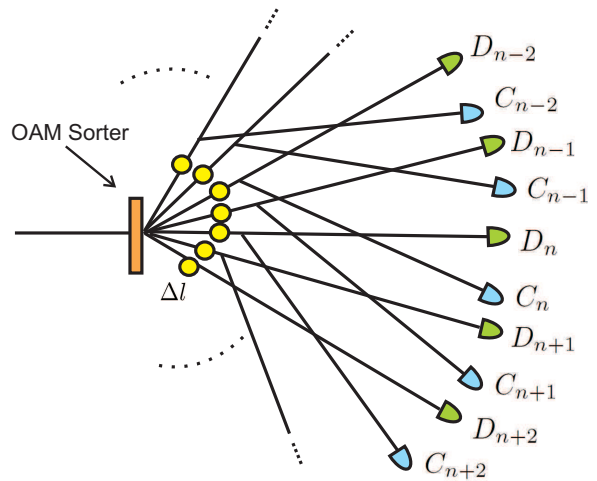}
\caption{Portion of a detector setup for discriminating Tribonacci states. Each yellow circle represents an OAM shifter that takes the OAM in that branch to zero.
Each place where lines split or cross is implied to have a beam splitter. The portion shown has OAM values from $l_n$ to $l_{n+4}$ coming out of the OAM
sorter.} \label{tribfig}
\end{center}
\end{figure}

The apparatus of Fig. \ref{CDfig} can be generalized in a number of ways. For example, consider the setup in Fig. \ref{tribfig}. Again, the OAM sorter sends
eigenstates outward in a series of spokes, but now a set of $50/50$ beam splitters causes each spoke to intersect with \emph{two} others. Although not drawn in,
it is implied that there are beam splitters at each of the points where lines in the diagram split or cross. Each of the yellow circles shifts the OAM to zero,
so that interference can occur. Suppose the lines going out from the beam splitter in the portion shown carry angular momenta $l_n, l_{n+1}, \dots, l_{n+4}$.
Assume that the state entering the sorter is a uniform superposition of these states $\sim \sum_n|l_n\rangle$. Then the state arriving at detector $D_n$ is
proportional to $|l_n\rangle -|l_{n-1}\rangle -|l_{n-2}\rangle $, while $C_n$ receives $i\left( |l_n\rangle +|l_{n-1}\rangle +|l_{n-2}\rangle\right) $. If
the states entering the OAM sorter are filtered to allow only a subset of the so-called Tribonacci numbers (obeying the three-term Tribonacci relation $T_n
=T_{n-1} +T_{n-2} +T_{n-3} $ \cite{feinberg,koshy}), then the setup, when fed with a source of entangled three-photon states, can be used to construct a three
photon protocol based on the Tribonacci numbers, in direct analogy to the two-photon Fibonacci protocol. Although rapidly becoming less practical with increasing
$N$, the extension to the $N-bonacci$ recurrence relation, $T_n= T_{n-1} + T_{n-2} + \dots T_{n-N} $ is obvious.

More generally still, a similar setup can be defined in which beam splitters cause each outgoing beam to intersect with any $n-1$ others before reaching a
detector, including intersections between beams that are not necessarily nearest neighbors. By further adding appropriate phase shifts and attenuation factors in
the beams, we may use the setup to detect other $n$-fold linear combinations of OAM states. Suppose, for example, that one wishes to test for linear combinations
of OAM values of the form $a_1|x_{m-1}\rangle +a_2|x_{m-2}\rangle + \dots +a_n|x_{m-n}\rangle $, where $\left\{x_m\right\}$ is some predetermined set of OAM
values. A similar recursive tree can be constructed such that: (i) Each outgoing line intersects $n$ beam splitters. (ii) Before the $j$th beam splitter in any
line, the intensity is attenuated by a factor of $t_j=|{{a_j}\over {a_{max}}}|$ (where $a_{max}$ is the largest of the coefficients) and phase shifted by $\phi
=arg(a_j)$. The result is that the $C$-type detectors will be sensitive to superposition states of the desired form. If the superpositions are non-orthogonal,
then there is of course still ambiguity in identifying them, but over multiple trials it will be possible to know whether the incoming states belong to this set
of superpositions or not by their statistics, as in section \ref{supersection}.

\begin{figure}
\begin{center}
\includegraphics[scale=.3]{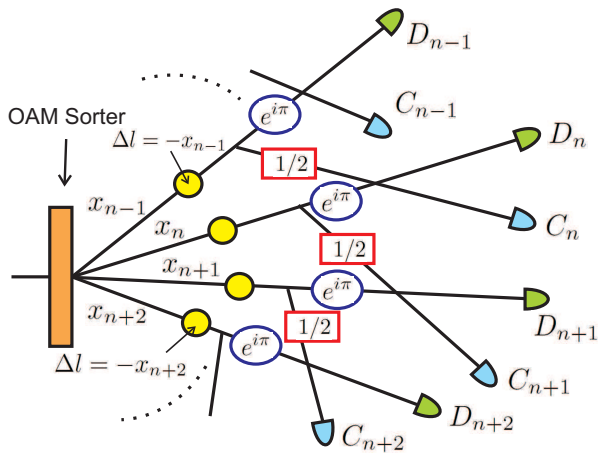}
\caption{(Color online) Portion of setup for detecting states of the form $-|x_{n}\rangle +{1\over 2}|x_{n-1}\rangle $, where $\left\{x_n\right\}$ is a predetermined set of
OAM values.
One output from the first beam splitter in the $n$th line is
attenuated in amplitude by ${1\over 2}$ and the other output is phase shifted by $\pi$. } \label{jumpfig}
\end{center}
\end{figure}

Any desired linear combination of OAM states can be detected in a similar manner. As one particular example, Fig. \ref{jumpfig} shows a portion of a setup in
which the $C_n$ detectors test for states of the form $-|x_n\rangle +{1\over 2}|x_{n-1}\rangle $, for some known set $\left\{x_n\right\}$ of OAM values. Again,
if the input is an OAM eigenstate with $l=x_n$, then the $C_n$ and $D_n$ detectors will fire with equal probability, while $C_{n+1}$ and $D_{n+1}$ will fire with one quarter of the probability. In contrast, if the input state is proportional to $-|x_n\rangle +{1\over 2}|x_{n-1}\rangle $, then the $D$ detectors will not fire due to destructive interference. The same setup, but with the beam splitters mixing next-nearest neighbors instead of nearest neighbor beams would
similarly test for states of the form $-|x_n\rangle +{1\over 2}|x_{n-2}\rangle$.

%


Further note that if the additional ingredient of coincidence counting is added, then sufficient numbers of detectors and beam splitters will allow detection of
various two-particle states in a similar manner. A two-photon bilinear state such as $|l_{n+4}\rangle \cdot \left( |l_n\rangle +|l_{n+1}\rangle
+|l_{n+2}\rangle\right)$, for example, can be detected by connecting a coincidence counter between the output of detector  $C_n$ in Fig. \ref{tribfig} and a detector
placed directly in the $l_{n+4}$ output of the OAM sorter.


\begin{figure}
\begin{center}
\subfigure[]{
\includegraphics[height=2.0in]{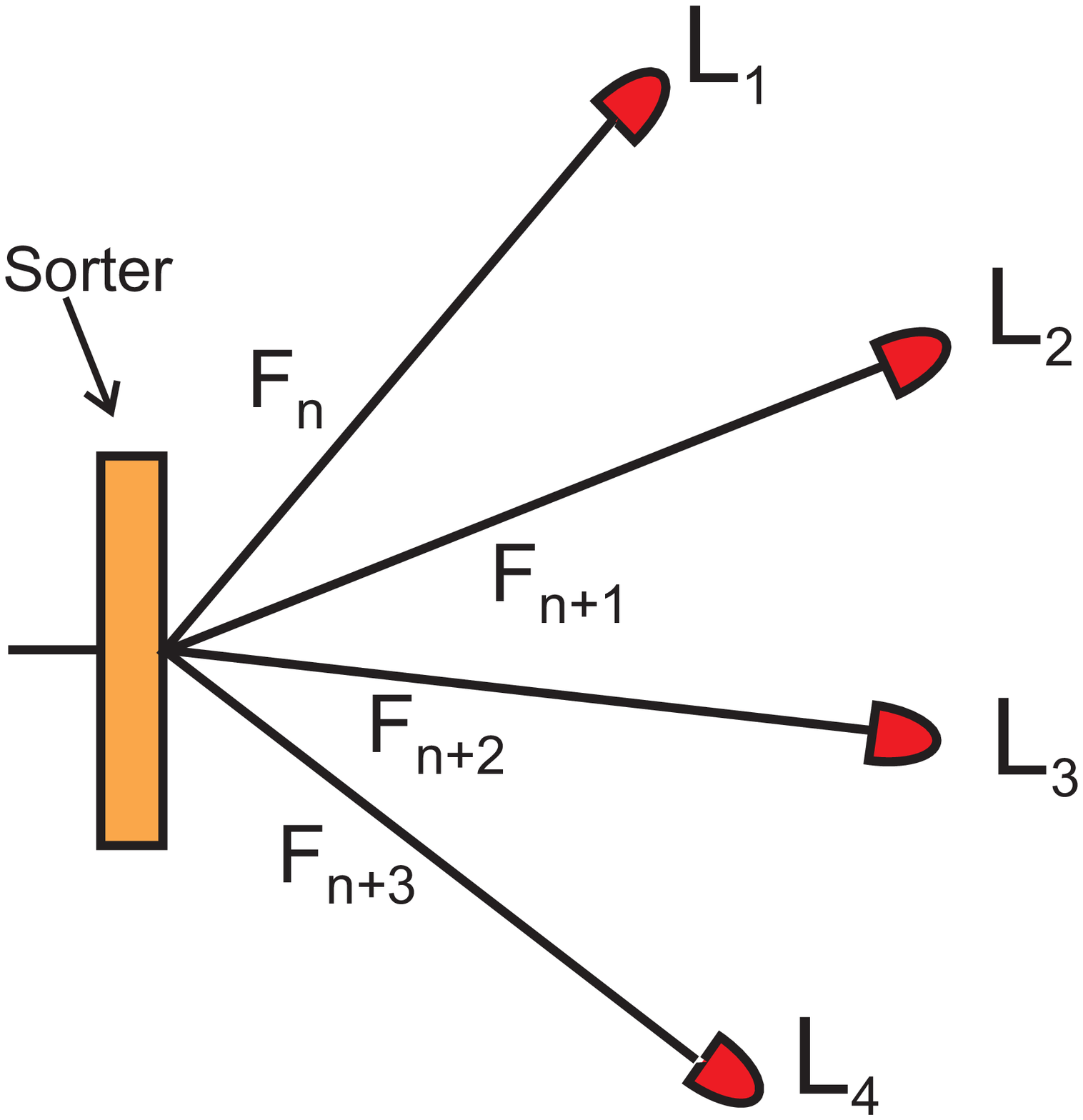}}
\subfigure[]{
\includegraphics[height=2.0in]{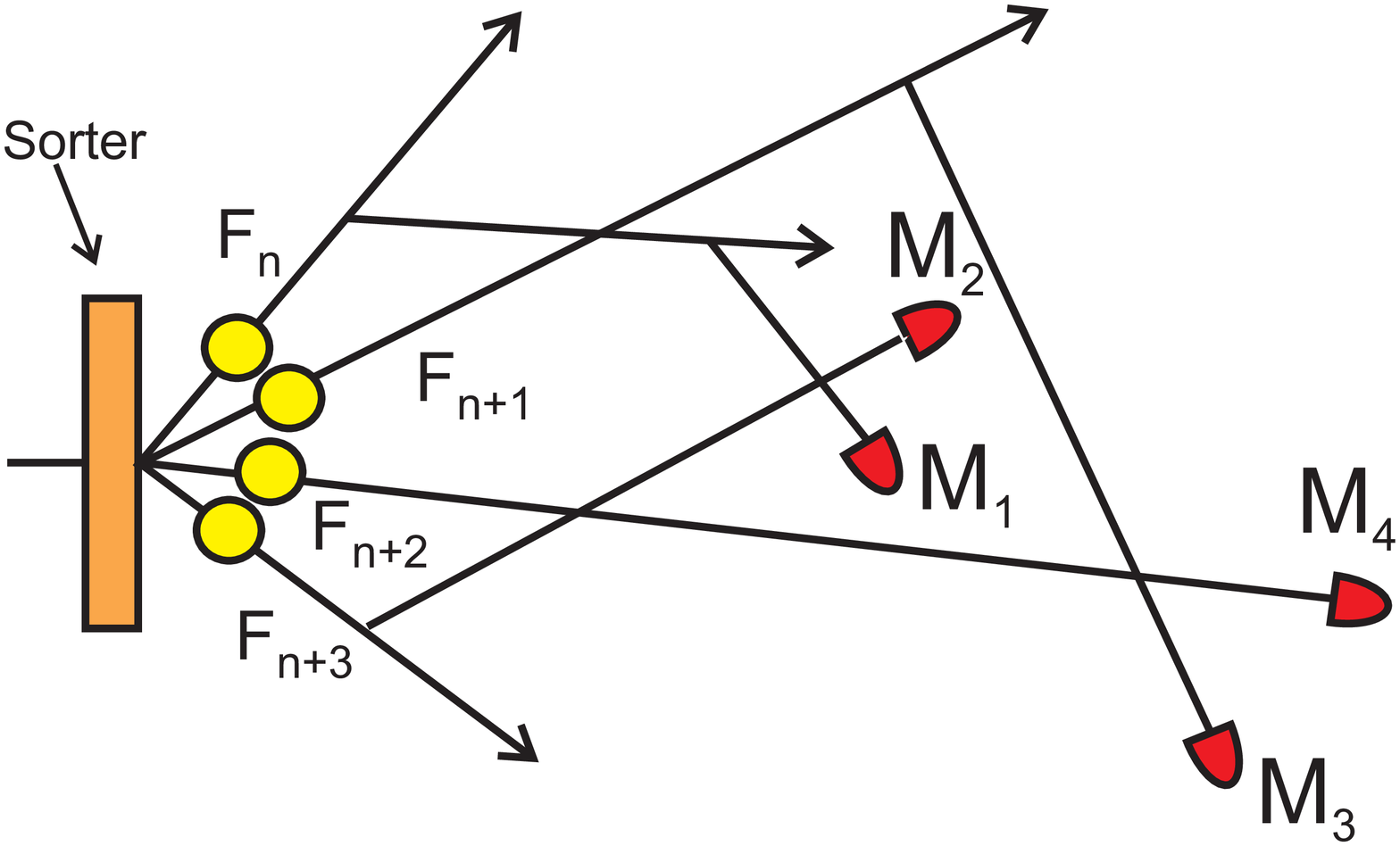}}
\caption{(Color online) (a) An OAM sorter followed by detectors makes measurements in a four-dimensional space spanned by the four incoming basis states (the states $|F_n\rangle,\dots |F_{m+3}\rangle$ in the case drawn). (b) An arrangement for measuring in a basis that is mutually unbiased to the basis of (a). The yellow circles shift the OAM to zero, so that interference can occur, and $50/50$ beam splitters are implied at each of the points where lines cross.  Additional beam splitters (not shown) are needed in some of the lines to equalize the detection probabilities in the four detectors ($M_1,\dots ,M_4$). }\label{eigoutcomesfig}
\label{mubfig}\end{center}
\end{figure}

One particular example of the usefulness of this approach can be seen by considering the switching between pairs of mutually orthogonal bases of dimension greater than two. In Fig. \ref{mubfig}, detection stages for a pair of mutually unbiased bases of dimension four are shown. Fig. \ref{mubfig} (a) shows the setup for detection in the ${\cal L}$ basis (OAM eigenbasis), for the four-dimensional space spanned by states $|F_n\rangle, |F_{n+1}\rangle ,|F_{n+2}\rangle, |F_{n+3}\rangle $. Fig. \ref{mubfig} (b) shows an arrangement that detects states in a basis that is mutually unbiased with respect to ${\cal L}$. In this second arrangement, the detectors $M_1,M_2,M_3,M_4$, respectively, detect the states \begin{eqnarray}|\psi_1\rangle &=&{1\over 2}\left\{ |F_n\rangle + |F_{n+1}\rangle +|F_{n+2}\rangle + |F_{n+3}\right\}\\ |\psi_2\rangle &=& {1\over 2}\left\{ |F_n\rangle + |F_{n+1}\rangle -|F_{n+2}\rangle - |F_{n+3}\right\}\\ |\psi_3\rangle &=& {1\over 2}\left\{ |F_n\rangle - |F_{n+1}\rangle -|F_{n+2}\rangle + |F_{n+3}\right\}\\ |\psi_4\rangle &=& {1\over 2}\left\{ |F_n\rangle -|F_{n+1}\rangle +|F_{n+2}\rangle - |F_{n+3}\right\}. \end{eqnarray}  Random switching between the two bases can be done passively, by having a beam splitter randomly send each photon to one arrangement or the other. This is in contrast to the setup in the original experiment of \cite{grob}, which switched between two mutually unbiased bases of dimension three by means of a complicated arrangement of mechanically shifted holograms.

\section{State synthesis}\label{stategensection}

We note that the situation presented in the previous sections contains two different sequences of linear combinations. There is the recursive sequence of values
$F_n=F_{n-1}+F_{n-2}$ on the space of real numbers, and there is the nonrecursive sequence of states $|F_n\rangle$ (or equivalently, of $|C_n\rangle $ and
$|D_n\rangle $) on the Hilbert space of the system. This leads to the question of whether similar optical arrangements to those shown above can produce more
general linear combinations on the Hilbert space, and whether we can in fact construct recursive sequences of states in Hilbert space. In general this would mean
a sequence of states in which each is proportional to a linear combination of several of the previous states in the sequence, according to some regular rule. We
require proportionality rather than equality due to the need to normalize the states.

With some slight alterations, the interferometric trees of the previous sections can be used to prepare desired superposition states from incoming OAM
eigenstates, instead of detecting different types of pre-existing superpositions. By removing the detectors and OAM shifters (the yellow circles in the figures
of the previous sections) and by illuminating the sorter with a uniform superposition of OAM states, $\sum_l|l\rangle$, we may arrange for different superpositions to appear at the
various output lines on the right. For example, the same arrangements described in the last section to detect states $ a_1 |l_{m-1}\rangle +a_2| l_{m-2}\rangle + \dots +a_n|l_{m-n}\rangle $ can, when detectors and OAM shifters are removed, \emph{prepare} these same states at the locations previously occupied by the $C$ detectors.

For example, consider Fig. \ref{CDfig} again, but now with these changes.
The exit ports where the $C$-type detectors previously were will now output states proportional to $|l_n\rangle+|l_{n-2}\rangle$. Similarly,
removing the $D$-type detectors, the corresponding output ports will produce states proportional to  $|l_n\rangle-|l_{n-2}\rangle$. By building up more
complicated trees, we can construct outputs that are arbitrary linear combinations of the OAM states. This ability to tailor different superpositions of OAM
states may be useful for a number of applications in quantum communication and quantum computing.

\begin{figure}
\begin{center}
\includegraphics[scale=.3]{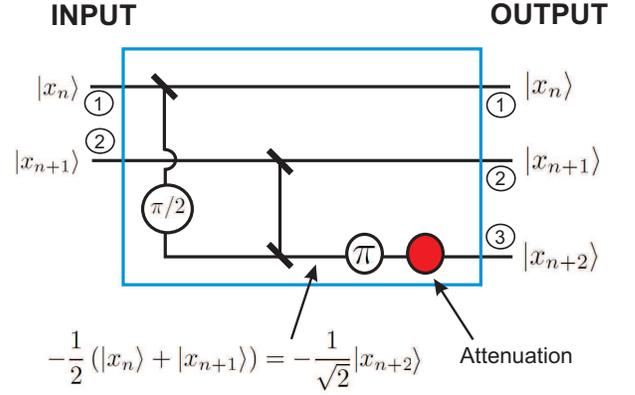}
\caption{(Color online) The basic unit cell for the recursive state generator (RSG) in the case of the Fibonacci recurrence relation.
Given a superposition of two input state (for example $|x_n\rangle =|F_n\rangle$ and $|x_{n+1}\rangle =|F_{n+1}\rangle$), the cell produces a state
proportional to $ |x_{n+2}\rangle \equiv {1\over \sqrt{2}}\left( |x_n\rangle +|x_{n+1}\rangle.\right) $. The attenuation of the amplitude in the bottom line by ${1\over \sqrt{2}}$ is there to ensure that the amplitudes of the three states are equal. By adding additional attenuations and phases shifts in the
various branches, units cells for any other complex, linear recursion relation can be constructed.} \label{unitfig}
\end{center}
\end{figure}

\begin{figure}
\begin{center}
\includegraphics[scale=.3]{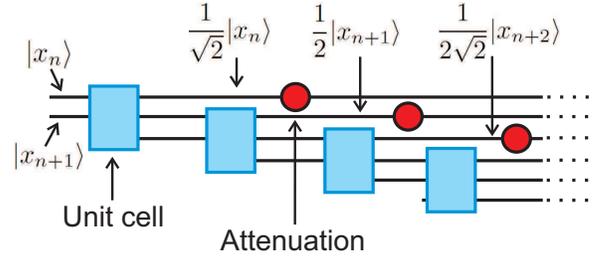}
\caption{(Color online) Recursive state generator (RSG).  Multiple copies of the unit cell of Fig. \ref{unitfig} (represented by the blue boxes)
can be concatenated in a fractal-like manner to form a set of output states that obey the Fibonacci relation
$|x_n\rangle ={1\over \sqrt{2}}|x_{n-1}\rangle +|x_{n-2}\rangle $. Each additional cell added has half the output intensity of the previous cell.
The attenuations (beam splitters), represented by the red circles, are added to make all of the final output intensities equal. Adding additional
attenuations and phases shifts inside the unit cell, other recurrence relations can be realized in Hilbert space. }\label{fractalfig}
\end{center}
\end{figure}

Finally, consider the possibility of creating states that themselves obey a recurrence relation. As a concrete example, consider again the Fibonacci relation.
But now we wish the states themeselves, not their OAM values, to obey the relation. In other words, we desire a set of states $|x_n\rangle$ satisfying
$|x_n\rangle \sim |x_{n-1}\rangle +|x_{n-2}\rangle$, where $\sim$ denotes equality up to normalization and the $x_n$ are the allowed OAM values of the states.
Suppose we again illuminate an OAM sorter with a uniform superposition of OAM states, $\sum_l|l\rangle$. We then take two of the output lines from the sorter
(say those for $l= x_n$ and $l=x_{n+1}$) and feed them into the unit cell shown in Fig. \ref{unitfig}. The input state to the unit cell is then ${1\over
\sqrt{2}}\left( |x_n\rangle_1 +|x_{n+1}\rangle_2\right)$, where the labels $1$ and $2$ refer to the two input ports. Then in addition to the original
superposition leaving from the top two output ports on the right of the cell, there is an equal amplitude for the superposition state $|x_{n+2}\rangle\equiv
{1\over \sqrt{2}}\left( |x_n\rangle +|x_{n+1}\rangle\right) $ to exit the lower port. In other words, the input state ${1\over \sqrt{2}}\left( |x_n\rangle_1  +
|x_{n+1}\rangle_2\right)$ is transformed into the output state ${1\over \sqrt{3}}\left( |x_n\rangle_1  +|x_{n+1}\rangle_2 +|x_{n+2}\rangle_3\right)$. Now suppose
we nest multiple copies of these unit cells together as in Fig. \ref{fractalfig}. If the attenuation factors are appropriately adjusted, the states at the output
ports (assuming $p$ unit cells are used) will be $|x_n\rangle_1, |x_{n+1}\rangle_2, |x_{n+2}\rangle_3\dots |x_{n+p+1}\rangle_{p+2}$, each appearing with equal
amplitude. We refer to this arrangement as a \emph{recursive state generator } (RSG). (Recursively constructed quantum states have been considered before
\cite{jaeger} from a different perspective.)

As the number of unit cells nested in the RSG increases, the output intensity decreases exponentially: each additional unit drops the output per output port by a
factor of two. But there is nothing in the setup that is especially sensitive to high intensities, so by using large input intensities it can be arranged to have
a sufficiently large number of cells to produce complicated sets of states while maintaining useable output levels.

\begin{figure}
\begin{center}
\includegraphics[scale=.3]{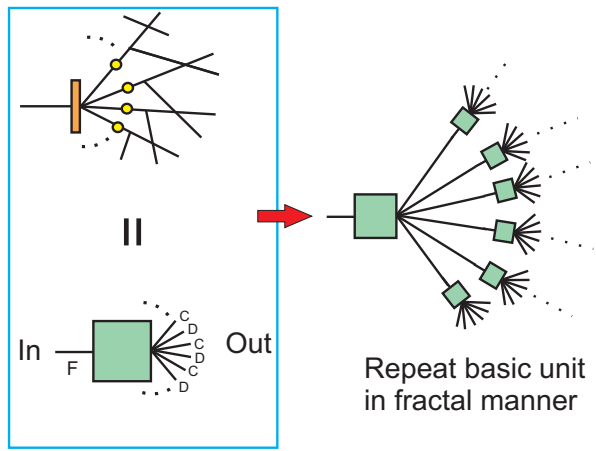}
\caption{(Color online) The apparatus of Fig. \ref{CDfig} (with the detectors removed) is abbreviated as a single unit, represented by the green box (left side of figure).
Multiple copies of this unit can be repeated, with the output of each used as input to the next (right side). The result is that the apparatus alternately
switches between the
$\mathbb{L}$ and $\mathbb{D}$ bases.} \label{walk1fig}
\end{center}
\end{figure}

\begin{figure}
\begin{center}
\includegraphics[scale=.3]{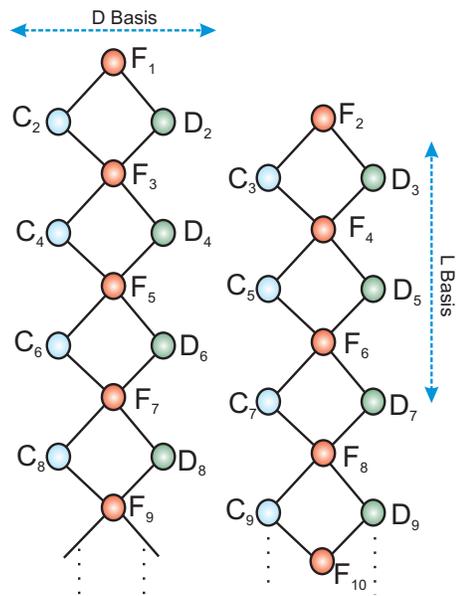}
\caption{(Color online) The apparatus of Fig. \ref{walk1fig} implements a random walk. By alternating $\mathbb{L}$ and $\mathbb{D}$ measurements, each of the two entangled photons can be made to walk randomly up and
down one of the two chains of states shown. The left-hand chain is constructed from states $|F_n\rangle$ with odd index $n$,
the right-hand chain is built from states with even $n$. The $|C_n\rangle$ and $|D_n\rangle$ states are those that are detected by the $C_n$ and $D_n$ detectors
in Fig. \ref{CDfig}.}\label{walk2fig}
\end{center}
\end{figure}

By placing the RSG at the output of an OAM sorter and allowing switching between different output lines of the sorter, different sequences obeying
the same recurrence relation are produced, for example switching between the Lucas and Fibonacci sequences.
More generally, by nesting multiple recursive trees inside each other, this gives access to finite approximations of fractal states in Hilbert space.
The construction of such
recursive chains of states allows the possibility of carrying out information processing or other tasks on these states in a relatively simple manner.

%

To give one application of the approach presented here, consider taking multiple copies of the apparatus from Fig. \ref{setupfig} and feeding the output of each
copy into the input of another copy in a fractal manner as shown in Fig. \ref{walk1fig}. If Alice and Bob both feed their half of a down conversion pair into
such an apparatus, the result is an implementation of an entangled two-photon quantum walk in OAM space: the OAM values of each photon can walk up and down the
two chains shown in Fig. \ref{walk2fig}. The OAM value can be thought of as the walk variable, with the interferometers that choose between the $C$ and $D$
states acting as the coin operators. (A quantum walk in OAM space has recently been implemented by other means in \cite{cardano}.)

\section{Conclusions}\label{conclusionsection}

In this paper, we have introduced an interferometric method for statistically distinguishing OAM superposition states both from eigenstates and from other
nonorthogonal superpositions. In addition, the same essential method can be used to\emph{ prepare} superpositions of optical OAM states.  This ability to tailor
different superpositions of OAM states may be useful for a number of applications in metrology and quantum computing; in particular, it allows complex sets of
states to prepared in a simple manner, allowing the possibility of carrying out sophisticated information processing algorithms directly on the Hilbert space of
the system. Further, by replacing the OAM sorters with devices that sort according to other degrees of freedom, similar arrangements can be used for superpositions
of eigenstates of other operators. The systems described can all be placed on integrated optical chips with current technology or with that which will be
available soon, allowing them to be used to generate or detect relatively complex Hilbert space structures in a simple and convenient manner.

\section*{Acknowledgements}
This research was supported by the DARPA QUINESS program through US Army Research Office award
W31P4Q-12-1-0015 and by the National Science Foundation under Grant No. ECCS-1309209.

\vfill

\end{document}